# Twist Angle Dependence of Exciton Resonances in $WSe_2/MoSe_2$ Moiré Heterostructures

**Chirag Chandrakant Palekar[1], Joakim Hagel[2], Barbara Rosa[1]*, Samuel Brem[3], Ching-Wen Shih[1], Imad Limame[1], Martin von Helversen[1], Sefaattin Tongay[4], Ermin Malic[3], Stephan Reitzenstein[1]****

[1]Institut für Festkörperphysik, Technische Universität Berlin, Hardenbergstrasse 36, 10623 Berlin, Germany

[2]Department of Physics, Chalmers University of Technology, 412 96 Gothenburg, Sweden

[3]Department of Physics, Philipps University Marburg, 35037 Marburg, Germany

[4]School for Engineering of Matter, Transport, and Energy, Arizona State University, Tempe, Arizona 85287, USA

* rosa@physik.tu.berlin.de, **stephan.reitzenstein@physik.tu-berlin.de

**Abstract:**

Van der Waals heterostructures based on TMDC semiconducting materials have emerged as promising materials due to their spin-valley properties efficiently contrived by the stacking-twist angle. The twist angle drastically alters the interlayer excitonic response by determining the spatial modulation, confining moiré potential, and atomic reconstruction in those systems. Nonetheless, the impact of the interlayer twist angle on the band alignment of the monolayers composing the heterostructure has received scant attention in the current research. Here, we systematically investigate the twist-angle dependence of intra- and interlayer excitons in twisted $WSe_2/MoSe_2$ heterobilayers. By performing photoluminescence excitation spectroscopy, we identify the twist-angle dependence of interlayer emission response, where an energy redshift of about 100 meV was observed for increasing twist angles. The applied microscopic theory predicts, on the contrary, a blueshift, which suggests that additional features, such as atomic reconstruction, may also surpass the moiré potential confinement. Those findings also prompt the effects of dielectric screening by addressing the redshift response to the stacking layer order. Furthermore, our findings support the evidence of a band offset dependence on the twist angle for the adjacent monolayers composing the heterobilayer system. Our fundamental study of exciton resonances deepens the current understanding of the physics of twisted TMDC heterostructures and paves the way for future experiments and theoretical works.

## Introduction

Transition metal dichalcogenides (TMDC) monolayers have first emerged as a novel class of semiconductors due to the appearance of a direct bandgap at two inequivalent ±K valleys when thinned down to monolayer (ML) of material [1,2]. Further unique physics, such as a strong spin-orbit coupling [3,4], and a lack of inversion symmetry promote the lifting of the spin degeneracy and the locking of the spin and valley degrees of freedom, which allows the ±K valleys to be individually addressed by circularly polarized light [5,6].

Excitonic properties of TMDCs appear even more attractive when monolayers are vertically stacked to form van der Waals (vdW) heterostructures (HSs). Among their novelties is the presence of a type-II band alignment in $MoSe_2/WSe_2$ HS which gives rise to spatially indirect interlayer excitons (IX) with, residing electrons and holes in conduction and valence bands of the adjacent layers. Moreover, the lattice mismatch and twist angle between the monolayers forms a moiré superlattice in the system, which creates a periodic potential landscape capable of trapping the interlayer excitons [7,8]. Those trapping effects, nonetheless, depend drastically on the twist angles that control the moiré periodicity and, therefore, their confining potential [9,10]. However, large twist angles leads to diminishing confining potential and delocalised excitons with altered optical response [9]. Additionally, it has been reported that a preferential stacking order of an asymmetric heterobilayer (e.g. $WSe_2/MoSe_2$/substrate and $MoSe_2/WSe_2$/substrate) alters the dielectric environment experienced by each constituent ML, and therefore, the optoelectronic response of the composed system [11]. Central to the physics of artificially stacked TMDC HS is the defined twist angle as well as the stacking order of the constitute layers, which appear as an effective and convenient tuning knob to control their optoelectronic properties [8,12,13]. Nonetheless, the influence of the twist angle on the band alignment of the monolayers composing the heterostructure has received scant attention in the recent research. Hence, twist angle dependent studies are highly desirable to gain detail insight into exciton resonances of the TMDC HSs.

In this report, we systematically study mechanically stacked twisted $WSe_2/MoSe_2$ heterobilayers with twist angles θ varying between $0^0$ and $60^0$, aiming to understand twist angle dependent emission properties and excitonic resonances. By performing micro-photoluminescence excitation (μPLE) spectroscopy on twisted $WSe_2/MoSe_2$ heterobilayers, we investigate the dependence of the twist angle on the intra- and interlayer excitonic resonances. Here, the PL response displays a drastic PL intensity reduction varying $\theta$ from $0^0$ to $30^0$, where a redshift of about 100 meV in the emission energy is observed. Further we measured blueshift in emission energy of IX for $\theta$ increasing towards $60^{0.}$ On the other hand, our microscopic theory suggests a blueshift in energy of interlayer excitons with increasing twist angle $\theta$ up to 10°, due to modulations in the moiré potential with decreasing supercell size and delocalization of interlayer exciton in the heterobilayer. This indicates that additional band gap variations, based on atomic reconstruction or dielectric screening effects, not captured by the applied model are responsible for the observable redshift in energy of interlayer excitons. Additionally, µPLE measurements demonstrate a noticeable twist-angle dependence of intralayer exciton (X) resonances of $WSe_2$ and $MoSe_2$, revealing evidence of the monolayer band alignment dependence with twist angle in a TMDC HS system. Overall, we provide a fundamental and systematic study of the TMDC heterostructures, which sheds light on intriguing effects of twist angle-assisted tuning and manipulation of the excitonic resonances.

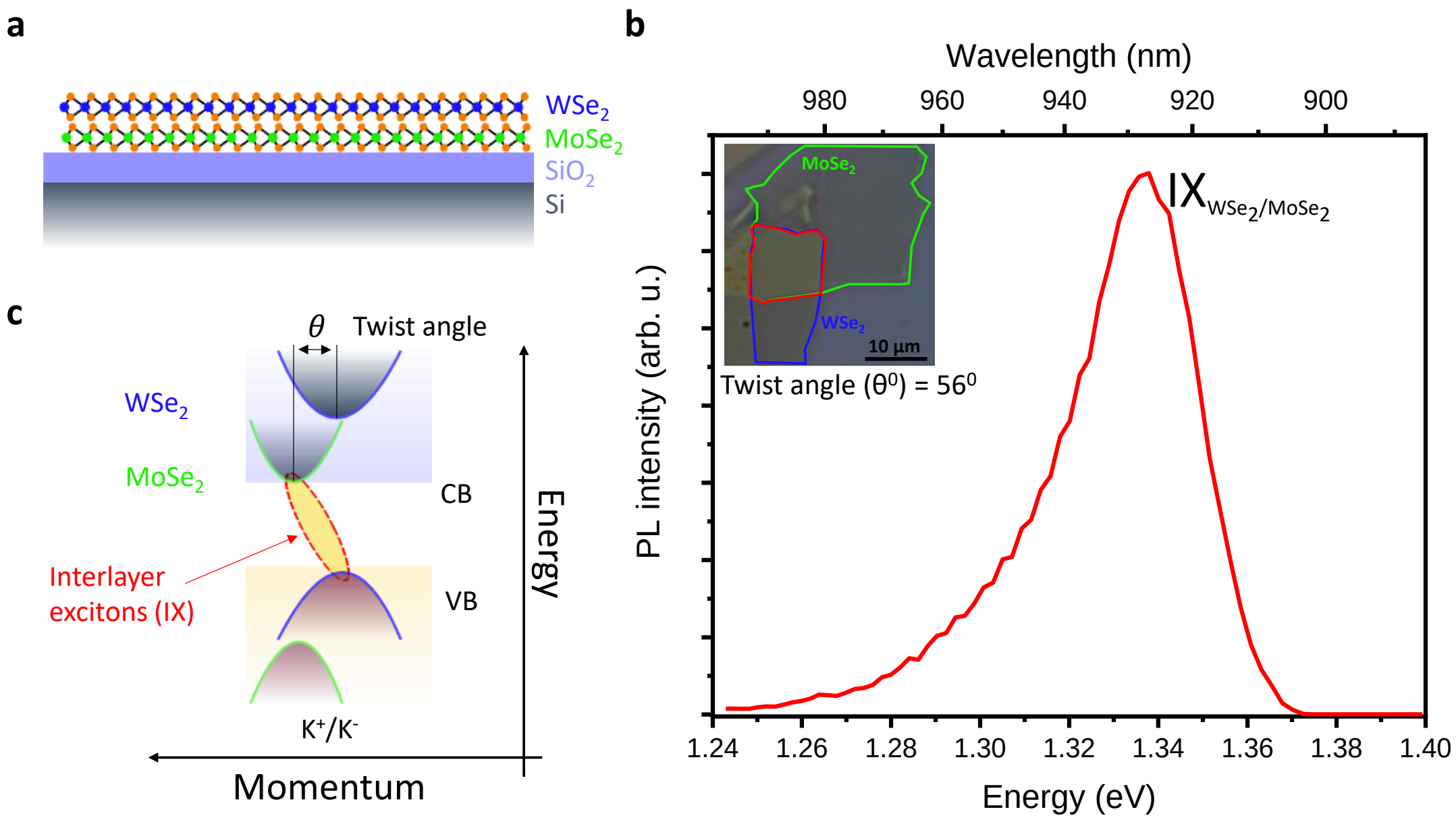

**Figure 1. Twisted $WSe_2/MoSe_2$ heterobilayer. a)** Schematic representing the stacking order of a $WSe_2/MoSe_2$ heterobilayer on $SiO_2/Si$ substrate. **b)** IX PL emission from a $WSe_2/MoSe_2$ heterobilayer with a twist angle of 56° at 4 K. The inset shows an optical micrograph of the fabricated twisted heterobilayer, highlighted with a red outline, consisting of $MoSe_2$ (green) and $WSe_2$ (blue). **c)** Illustration of conduction and valence band configuration around K point in Brillouin zone with momentum mismatch as result of twist angle ($\theta$).

## Sample Fabrication

The $WSe_2/MoSe_2$ heterostructures were fabricated by employing the mechanical exfoliation [14] and dry-transfer method [15] techniques. Using a suitable low adhesive tape, the TMDC crystals are thinned and later exfoliated on a PMMA gel strip to assist the dry transfer technique. Monolayers of TMDC materials were thus identified by optical microscope images. Suitable $WSe_2$ and $MoSe_2$ monolayers were selected and aligned for the target twist angle considering the edges of each layer. Lastly, the monolayers were transferred onto a $SiO_2$ substrate, keeping the substrate temperature around 60 $^0$C. Figure 1a shows a schematic of the $WSe_2/MoSe_2$ HS system, in which the stacking order of the constituent monolayers is highlighted.

## Results

In this work, we performed room temperature second harmonic generation (SHG) measurements to determine the twist angle between the MLs, and micro-photoluminescence (PL) to study the twist angle dependence of the interlayer exciton emission energy. All PL measurements are carried out at 4K unless mentioned otherwise. Throughout our systematic research, we fabricated six high-quality $WSe_2/MoSe_2$ HSs with distinct twist angles varying from R-type stacking (close to 0°) to H-type stacking (close to 60°). More fabrication details are given in Methods section (Sample fabrication) and in the Supplementary Material, Fig. S1.

Figure 1b shows the representative $WSe_2/MoSe_2$ interlayer exciton photoluminescence extracted from the sample with $\theta$ = 56°, where one observes the emission peak at ≈ 1.33 eV, likewise previous works reporting similar heterostructures [13,16–18]. Aiming to achieve the highest photoluminescence quantum yield, we excited the sample in resonance with a $WSe_2$ neutral exciton emission energy $E_{exc}$ = 1.71 eV. As stated in the literature, the twist angle between stacked monolayers rules the momentum mismatch around the ±K valleys in the Brillouin zone, which promotes the formation of momentum direct (bright) or momentum indirect (dark) interlayer excitons (dark exciton in HBLs) [19–21]. At 0° and 60° twist angle, i.e. R-type stacking or H-type stacking, respectively, the type-II band alignment of the two adjacent monolayers holds the minimum momentum mismatch at ±K points. In contrast, with the increase (decrease) of the twist angle from $0^0$ ($60^0$) into the largest angle misalignment ($\theta \approx 30^0$), the transition becomes more indirect as a result of increasing momentum-space mismatch [22,23]. Figure 1c depicts the two scenarios described above.

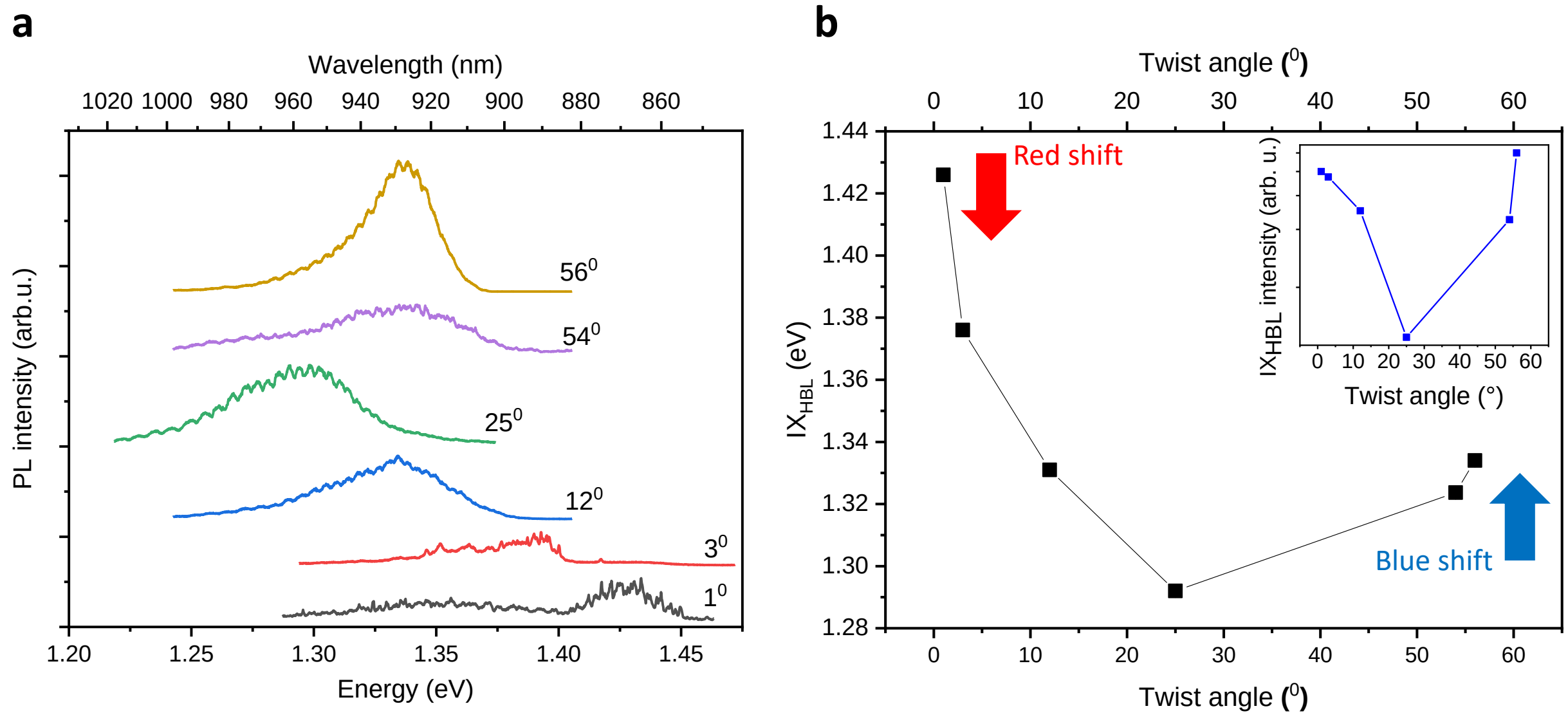

**Figure 2. Twist angle dependence of interlayer exciton emission in $WSe_2/MoSe_2$ heterobilayers. a)** Interlayer exciton PL emission from twisted $WSe_2/MoSe_2$ heterobilayers with different twist angles. **b)** IX emission energy as function of twist angle exhibiting the substantial redshift from 0° to 25°. Inset: IX intensity as function of twist angle with a pronounced minimum at an angle of 25°. The interlayer exciton PL intensity drastically reduces with increasing (decreasing) twist angle from $0^0$ ($60^0$).

## Twist angle dependence

It is well established that the lattice mismatch in the HS system and twist angle leads to the formation of high or low symmetry stacking orientations. Such alignments actively alters the lateral interface between the $WSe_2$ and $MoSe_2$ monolayers by modulating the physical interlayer separation (in real space) and the local atomic registry as a consequence of respective layers alignment [8,9]. Therefore, the local atomic registry depends on the twist angle alignment, which modulates the interlayer coupling conditions. Those variations can be noticed through optical properties of IX, such as PL intensity or emission energy. To obtain a better overview of the evolution of IX emission for different twist angles, we illustrate the normalized PL response of our samples in Fig. 2a.

Figure 2b shows the extracted IX emission energy as a function of twist angle, in which we observe IX energy manifesting a systematic redshift to blueshift (positive parabolic appearance) on the order of 100 meV as the angle changes from $0^o$ to $60^o$. Here, we notice that the PL intensity of the IX with a twist angle close to $0^0$ or $60^0$ is a maximum, whereas it is drastically lower for the intermediate angles (see Fig. 2b inset). Moreover, it is worth mentioning that relatively longer exposure times (10 times longer) and higher excitation powers (more than twice) were used to get a reasonable signal-to-noise ratio in PL response for HSs with twist angles of 12° and 25°. This PL intensity reduction can be understood as a weakening of interlayer coupling strength for intermediate twist angles [23]. In fact, the interlayer separation between the MLs gradually increases with twist angle (up to $30^0$), and so the associated interlayer coupling strength depletes gradually [24]. Consequently, the reduced the IX exciton population leads to weakened emission intensity as observed for $\theta = 25^o$ (Fig. 2b, inset). On the other hand, high symmetric stacking and minimal interlayer separation result in higher interlayer coupling strength at twist angles close to $0^0$ and $60^0$, which promotes the efficient charge transfer between the $WSe_2$/$MoSe_2$ interfaces [23].

To obtain a better understanding of our experimental findings, the energetic positions of the IX have been modeled with a microscopic theory, distinguishing between different interlayer coupling mechanisms that can give rise to twist angle-dependent spectral shifts of the IX resonance [9,25–27]. Microscopic access to the moiré exciton energy landscape is gained by describing the periodic moiré potential as a modification of the decoupled monolayer energies, which can be solved from the Wannier equation [9,26–28]. These modifications include stacking-dependent alignment shifts and electron/hole tunneling. The resulting spectral shift is due to an electrostatic potential [29], giving rise to a renormalization of the band structure [9,25,28], and the tunneling is due to the interlayer wavefunction overlap [26,28]. Since the exciton resonance in question is assigned to the bright K-K exciton, the tunneling of carriers is weak, and thus the alignment shift plays the crucial role [25,28]. Our calculations predict a blueshift of about 40 meV for the low-lying bright interlayer exciton, stemming from the decrease in the supercell size, which suppresses the impact of the alignment shift on the exciton, consequently delocalizing the exciton in real space (see Supplementary Material, section 5). Interestingly, the theory results contradict the experimentally observed redshift (see figure 2b). This discrepancy could be due to distinct neglected effects, as atomic reconstruction, which in the low twist angle regime strongly affects the exciton energy landscape [30,31].

Nonetheless, compared to our experimental results, it was reported that a $MoSe_2/WSe_2$ heterostructure exposed to different top and bottom environments might experience a dielectric screening asymmetry and affect the optical response of interlayer excitons [32–34]. It has observed that transferring $MoSe_2$ onto $WSe_2$ leads to a blueshift -- redshift (negative parabolic appearance) of PL response as a function of twist angle [22], this also in agreement with our theoretical predictions. On the other hand, as observed in our work (see figure 2) and discussed in the Supplementary Material of Ref. [29], the samples were stacked in the reversed order – the $WSe_2$ ML was placed onto $MoSe_2$ ML—, exhibit, on the contrary, a considerable redshift – blueshift (positive parabolic appearance) dependent on the twist angle.

The overall observations suggest that the stacking order of the constituent monolayers plays a substantial part in the twist-angle dependence of the IX emission energies. Importantly, the redshift in IX emission is also observed for the HSs stacked on few layers of hBN (for details see Supplementary Material, section 4). Such a scenario provides additional freedom to control the IX emission in twisted HS systems along with the deterministic twist angle. However, a straightforward understanding is still desired, and could potentially be achieved through spectroscopy measurements combined with further experiments, such as scanning probe microscopy techniques.

## Band offset modulation based on twist angle

Aiming at a deeper understanding of our HS system, we performed twist-angle dependent µPLE measurements on all samples mentioned in Fig. 2a. The PLE signal of IX emission from the HS was recorded as a function of excitation energy, ranging from 1.75 to 1.56 eV and under constant excitation laser power. Figure 3a shows a 2D false color PLE map of IX emission, which features two distinct and characteristic resonances for the HS system with a twist angle of 56°. The PLE resonances are associated with intralayer excitons of the constituent monolayers, $X_{WSe_2}$ and $X_{MoSe_2}$. The relative integrated IX intensity as a function of excitation energy is displayed in Fig. 3b, in which $X_{WSe_2}$and $X_{MoSe_2}$ resonances were extracted by fitting the data with a Gaussian equation.

Besides the widely reported twist angle dependent interlayer emission response of HSs, it is expected that the band structure of each monolayer will also be affected by the angle between the layers, likely due to the proximity effects, already mentioned in the section before and in [32], [33], [34]. Considering those aspects and the fact that PLE data gives information about the absorption of the individual layers and, therefore, the constituent band structures, we decided to investigate the intralayer response as a function of the twist angle of $MoSe_2$ and $WSe_2$ composing the HS. The resulting evolution of exciton resonance dependent on θ is displayed in Fig. 4a. We first noticed that both resonances, $X_{WSe_2}$ and $X_{MoSe_2}$, exhibit a negative parabolic appearance with apparent blueshift. Specifically, the $X_{WSe_2}$ exhibits a blueshift up to

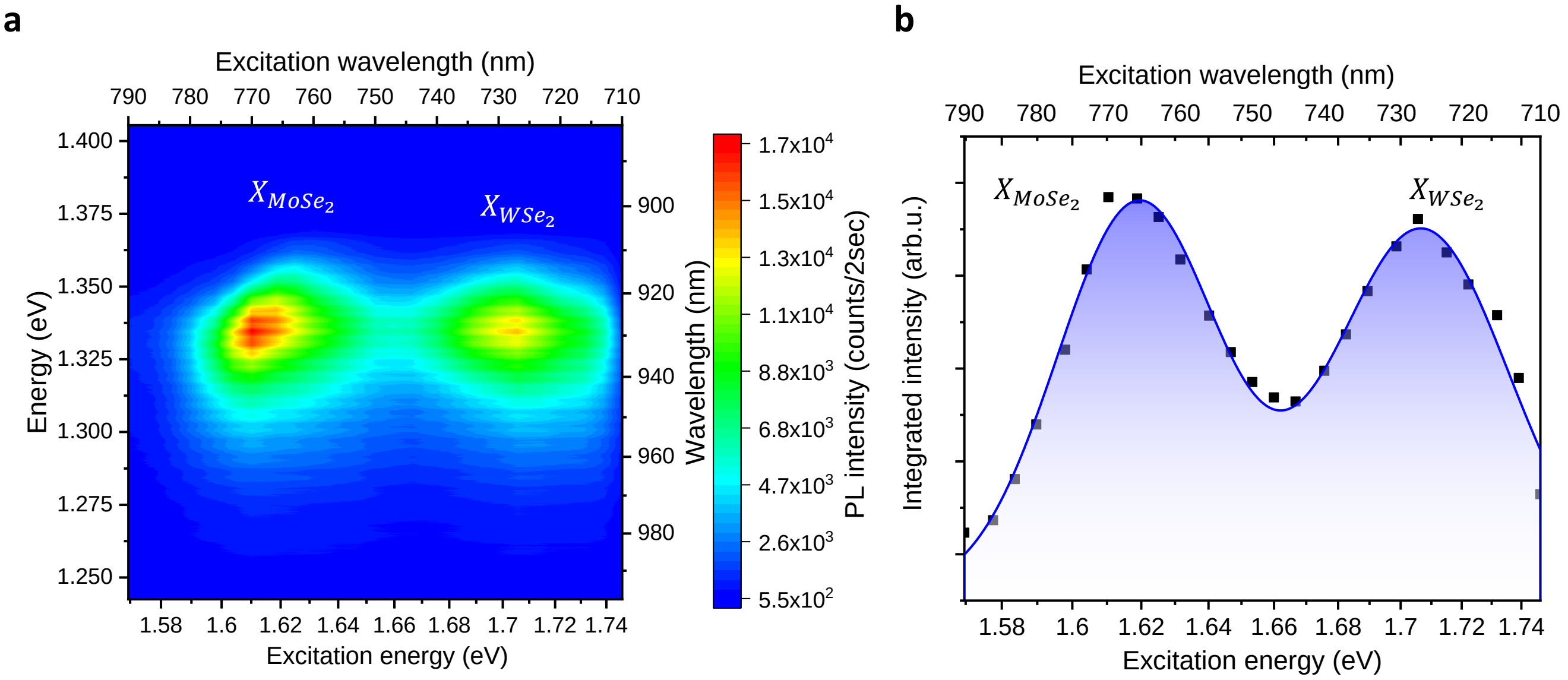

**Figure 3. Micro-photoluminescence excitation (µPLE) measurements on a $WSe_2$/$MoSe_2$ heterobilayer with a twist angle of $25^0$. a)** IX emission as a function of excitation energy (wavelength) under constant excitation power. **b)** Integrated intensity (black square) as function of the excitation wavelength exhibiting resonances associated with intralayer exciton. The integrated intensity is fitted with Gaussian function (blue line).

34 meV over the span of twist angles from $0^0$ to $25^0$. For twist angles close to $60^0$, the resonances display a pronounced redshift. It positively confirms that the intrinsic intralayer exciton resonances also undergo a modification of their optical properties as a function of twist angle. One cannot exclude other external effects, such as dielectric screening or local defects and strain, which may interfere in the Coulomb interaction of excitons. However, the increment observed through PLE resonance separations with increasing twist angle is consistent even in $WSe_2/MoSe_2$ HSs stacked on $hBN/SiO_2$ substrate (see Supplementary Material Fig. S5).

Our findings may also reveal evidence of the monolayer band alignment dependence with twist angle in a TMDC HS system. It is well established that the band offset of conventional semiconductors is a key parameter for optoelectronic devices design because it controls the carrier occupation and, thus, the transport phenomena [35]. Nevertheless, the band offset can be manipulated by altering the Fermi level of the individual materials layers at thermal equilibrium by, for instance, by varying the local density of carriers (e.g. electrostatic doping) [36–40]. In the TMDC family, however, further aspects must be considered when studying a heterojunction formed by few- or monolayers. Few theoretical and experimental reports have discussed the band alignment of different monolayer materials [34] [40], and its importance in the 2D materials heterostructure properties. However, the mechanism by which not only the choice of materials but also the angle between the layers forming the junction affects the band offset of those systems have not been yet established.

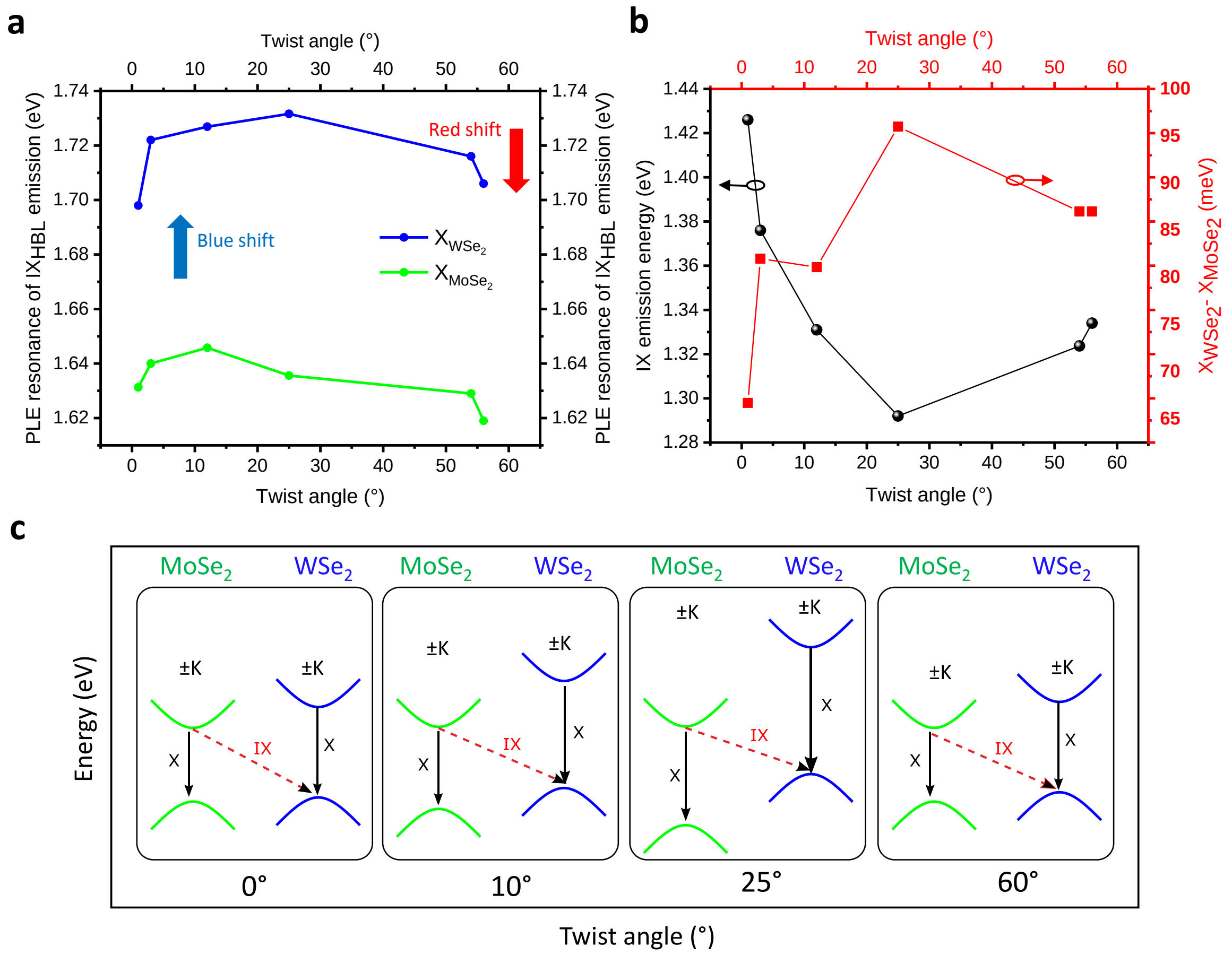

**Figure 4. Influence of twist angle on PLE resonance energies of a TMDC HS. a)** PLE resonance energy of and for HS $WSe_2/MoSe_2$ with different twist angles. **b)** Shows the inverse relationship between the PLE resonance separation as function of twist angle and with the IX emission energy. **C)** Schematic representation of the twist angle dependent change in band offset for twist angles 0°, 10°, 25° and 60°.

To better understand the potential correlation between the band offset and twist angles in our systems, we present in Fig. 4b the two data sets corresponding to the interlayer and intralayer resonances. The blue graph in Fig. 4b shows to the PLE resonance separation $\left(X_{WSe_2} - X_{MoSe_2}\right)$ as a function of the twist angle extracted from the PLE data of Fig. 4a. In comparison, the red graph depicts emission energy of IX (the data already presented in Fig. 2b). We first note that both exciton complexes (intra- and inter-layer) respond inversely to the twist angle, which might indicate that after the stacking, the junction of conduction and valence bands of the constituent materials, forming the type-II alignment, arranges differently for each twist angle. The bandgap of TMDCs is also sensitive to their thickness [34,41], which suggests that the proximity of layers, already defined as deeply dependent on twist angle, may interfere in the band alignment of TMDC heterostructures as well. This scheme is seen in Fig. 4c. In the transitions lying in ±K valleys of the materials, the intralayer exciton responds inversely to the interlayer exciton dependent on the twist angle. Our understanding based on our theoretical and experimental outcomes might not explain the exact junction formation of heterostructures in distinct dielectric screenings; nonetheless, it sheds light on the band offset dependence on the materials composing a heterobilayer as much as the twist angle between the layers is concern.

## Conclusion

In this work, we fabricated high-quality twisted heterostructures consisting of $WSe_2$ and $MoSe_2$ monolayers to investigate the influence of twist angle on the intra- and inter-layer excitonic properties. The studied heterostructure samples have twist angles ranging from $0^0$ to $56^0$, and feature pronounced interlayer exciton emission at cryogenic temperature. The interlayer exciton features a redshift in emission energy for twist angles ranging from $0^0$ and $25^0$. Based on modeled microscopic theory considering the moiré potential effect, the interlayer exciton emission shows a pronounced blueshift. The experimental redshift can be attributed to the proximity effect of constituent monolayers stacked in a specific order ($WSe_2/MoSe_2$). Furthermore, the photoluminescence excitation resonances also change systematically with the twist angle, which indicates a substantial influence of twist angle on the bandgap of individual materials. The separation between the $WSe_2$ and $MoSe_2$ resonances also differs with a change of twist angle, suggesting an alteration in the coupling strength between layers. This observation ultimately indicates changes in the band offset, which corresponds to the optically active type-II interlayer exciton transition, but also supports the experimentally determined reduced emission energy at twist angles leading to 30°. Taken together, our results reveal that the TMDC heterostructure's band gap, and therefore, the excitonic response, strongly depend not only on the symmetric atomic arrangements and confining moiré potentials but also on the dielectric environment and twist angle.

## ACKNOWLEDGEMENT:

Financial support by the Deutsche Forschungsgemeinschaft (DFG) by project Re2974/26-1 is gratefully acknowledged. The Marburg group acknowledges support from the DFG via SFB 1083 and the regular project 512604469.

## CONFLICT OF INTEREST:

The authors declare no conflict of interest.